# Advances in Classifying the Stages of Diabetic Retinopathy Using Convolutional Neural Networks in Low Memory Edge Devices

Aditya Jyoti Paul [1,2]

1. Cognitive Applications Research Lab,
India.

2. Department of Computer Science and Engineering,
SRM Institute of Science and Technology,
Kattankulathur, Tamil Nadu – 603203, India.
aditya_jyoti@srmuniv.edu.in

*Abstract* — Diabetic Retinopathy (DR) is a severe complication that may lead to retinal vascular damage and is one of the leading causes of vision impairment and blindness. DR broadly is classified into two stages – non-proliferative (NPDR), where there are almost no symptoms, except a few microaneurysms, and proliferative (PDR) involving a huge number of microaneurysms and hemorrhages, soft and hard exudates, neo-vascularization, macular ischemia or a combination of these, making it easier to detect. More specifically, DR is usually classified into five levels, labeled 0-4, from 0 indicating no DR to 4 which is most severe. This paper firstly presents a discussion on the risk factors of the disease, then surveys the recent literature on the topic followed by examining certain techniques which were found to be highly effective in improving the prognosis accuracy. Finally, a convolutional neural network model is proposed to detect all the stages of DR on a low-memory edge microcontroller. The model has a size of just 5.9 MB, accuracy and F1 score both of 94% and an inference speed of about 20 frames per second.

## I. INTRODUCTION

Diabetic Retinopathy (DR) is a malady affecting about a third of the estimated 285 million patients suffering from diabetes mellitus (DM), worldwide, and a third of the patients with DR suffer from the more severe vision-threatening DR, involving diabetic macular edema (DME) [1]. These numbers are reaching epidemic levels, with previous estimates of around 439 million DM patients by 2030 [2] and 600 million by 2040, skyrocketing to more recent estimates of 578 million (10.2%) by 2030 and a whopping 700 million (10.9%) by 2045, according to the diabetes Atlas 2019 [3]. About a third of these diabetes patients are expected to have Diabetic Retinopathy [4]. In India, the number of diabetes patients increased from 26 million in 1990 to 65 million in 2016 [5] and is fast increasing. India ranked second internationally with 77 million diabetes patients in 2019 (95% confidence interval between 62.4 - 96.4 million), with projections for 2045 skyrocketing to 134.2 million (95% confidence interval between 108.5 - 165.7), for the age group of 22-79 years [3].

The current COVID-19 pandemic has cast a disastrous cloud on the health and well-being of people around the globe [6]; in one survey, diabetes was found in about 47% of the hospitalized patients and was associated with severe affliction and mortality [7]. Thus, diabetes and its complication Diabetic Retinopathy pose significant challenges in today's world.

This research work will first survey the existing literature, critically analyzing the advances at the cross-section of bio-medical and AI research and then propose a novel Convolutional Neural Network based technique. Additionally, some pre-processing caveats would be discussed in detail in this paper which serves as the secret sauce to making a simple model gain such impressive performance metrics.

The rest of the paper is presented as follows: Section II discusses some preliminaries about Diabetic Retinopathy and the image processing and machine learning foundations necessary to understand and delve into the problem, Section III surveys the existing literature, Section IV presents the proposed algorithm, Section V presents and analyses the observations and results, Section VI serves as a conclusion, and throws light on some future research avenues in AI-aided real-time Diabetic Retinopathy prognosis.

## II. PRELIMINARIES

### A. Stages of Diabetic Retinopathy

Diabetic Retinopathy is broadly categorized into non-proliferative and proliferative and is further classified into 4 stages usually. The classifications of the disease can be more specific with more than four stages as well, as suggested in some literature, but for the purposes of this paper, only the four standard stages will be delved into, and the characteristics of those four stages are discussed below.

Stage 1 is mild non-proliferative Diabetic Retinopathy, which is the earliest stage, which involves tiny swellings in the retinal blood vessels, termed as microaneurysms. (MA) Minute fluid leakage into the retina, leading to the macula starting to swell may be observed, which would indicate the condition degenerating to Stage 2.

Stage 2 is moderate non-proliferative Diabetic Retinopathy, characterized by an increase in swelling of the tiny blood vessels, interfering with the blood flow to the retina, which would lead to a lack of proper nourishment. Blood and other fluids start accumulating in the macula and at least one retinal hemorrhage or MA is observed, possibly along with the presence of at least one of soft exudates (cotton-wool spots), hard exudates, venous beading.

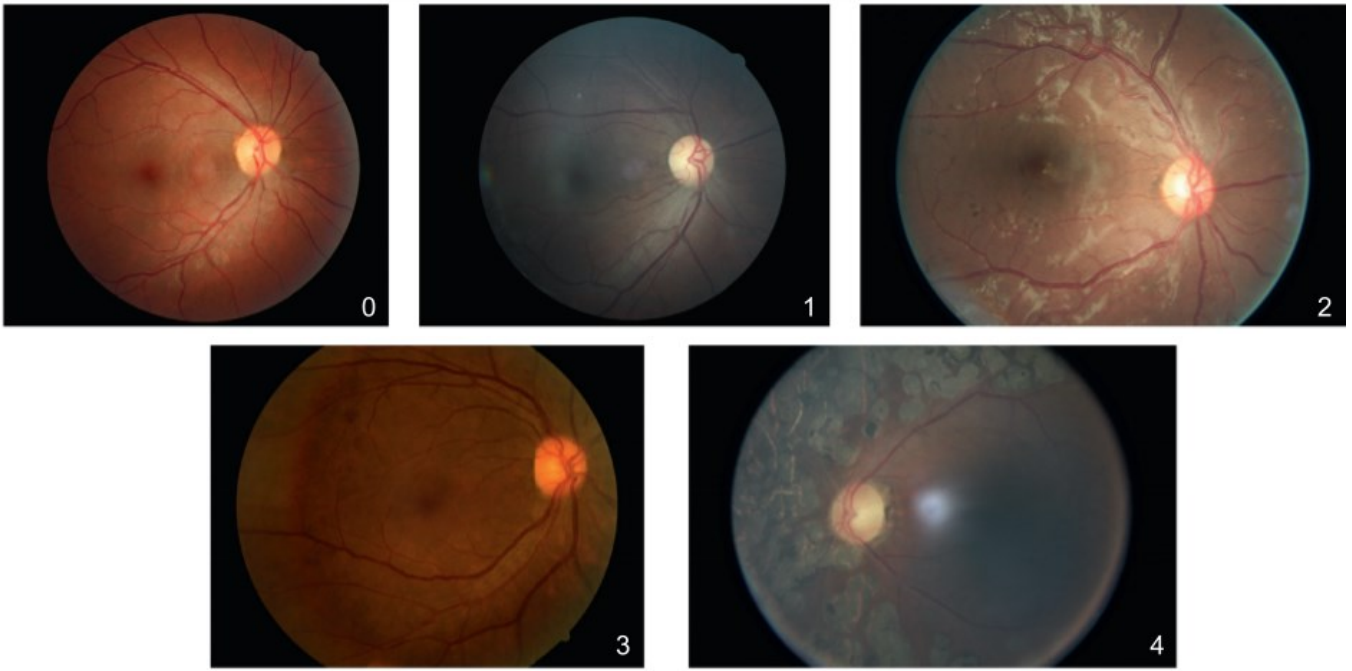

Fig. 1. Stages 0-4 of Diabetic retinopathy, with the labels marked in the corner of each image, adapted from [8].

Stage 3 is severe non-proliferative Diabetic Retinopathy, which involves a larger section of blocked blood vessels and the body receiving signals to grow new blood vessels in the retina. It is usually characterized by the "4-2-1 rule", i.e. hemorrhages and MAs or both are present in all the four retinal quadrants, venous beading appears in two or more retinal quadrants or intraretinal microvascular abnormalities (IRMA) are observed in at least one retinal quadrant. The presence of two or more of these criteria leads to 'very severe NPDR'.

Stage 4 is proliferative Diabetic Retinopathy (PDR). This is the most advanced stage of the condition and is characterized by neovascularization (NV), wherein new blood vessels start forming either on or within a disc diameter of the optic disc (NVD), or anywhere else on the retina (NVE). It would also be sufficient to observe a pre-retinal hemorrhage (PRH) or a vitreous hemorrhage (VH) to identify Stage 4 DR or PDR.

Another issue that often appears in diabetes patients is diabetic macular edema (DME), where intraretinal fluid accumulates in the macula, which also has a debilitating effect on visual acuity, leading to blindness, if left untreated.

A brief synopsis of the specifics of diagnosing DR can be found in [9, 10], and a more detailed discussion and insights can be gleaned from these works [11-16]. Fig. 1 shows a clear visual representation of the stages of diabetic retinopathy, adapted from a figure in [8].

*B. Retinal Imaging*

The technological advances in the field of retinal imaging have seen massive leaps in the past 170 years, since the invention of the first ophthalmoscope by Hermann Von Helmholtz. Currently, the two most common forms of non-invasive imaging are Fundus Imaging and Optical Coherence Tomography (OCT). Fluorescein Angiography is another procedure of visualizing the blood vessels in the eye, used in DR prognosis, which involves injecting a fluorescent die into the bloodstream, and it is an invasive procedure. The next two paragraphs discuss OCT and fundus imaging in more detail.

OCT works on the principle of backscatter, the depth at which a particular backscatter originates can be inferred from its time of flight. They originate mostly at the interfaces due to the refractive index differences between two tissues and thus the backscatter from deeper tissues can be distinguished from that caused by more superficial ones. Since the retinal thickness ranges between 300-500 μm, these differences are very minute and are measured by interferometry, specifically, low-coherent light interferometry, where the wavelengths are slightly longer than visible light.

Fundus imaging is a relatively broad modality of imaging retinas, involving various techniques. In brief, it can be defined as a modality projecting a two-dimensional representation of the three-dimensional diaphanous tissues onto the imaging plane, where the intensities represent the amount of reflected light. This can include, just a single wavelength band which is simple fundus photography, or all three R, G, B channels as in color fundus photography. Depth resolution could be achieved by viewing the retina from different angles, as carried out in stereo fundus photography, multiple wavelength bands could be used which is called hyperspectral imaging. Scanning Laser Ophthalmoscopy (SLO) involves time-sequenced image intensities representing the quantum of reflected confocal laser light, and improving upon it by optical correction of the wavelength aberrations by mathematical modeling is called adaptive optics SLO. Fluorescein and Indocyanine angiography come under the fundus imaging modality as they are two-dimensional as well.

It has been shown that color fundus imagery alone is often sufficient to diagnose diabetic retinopathy and this research and dataset focuses on that class of imagery, more details about the dataset would be provided in Section IV, after a detailed literature review analyzing the advances in diagnosing Diabetic Retinopathy in Section III.

## III. LITERATURE REVIEW

Diabetic Retinopathy is a disease affecting millions across the world and is the most common cause of blindness in people under the age of fifty [17]. Machine Learning and Computer Vision assisted diagnostic techniques have been playing a major role in screening patients worldwide, reaching out to places where skilled doctors and ophthalmologists are not available. This section discusses the advances made to date in this field, building up from the early diabetic retinopathy research to the current state-of-the-art techniques in diabetic retinopathy diagnosis using various automated techniques, besides throwing light on the imaging and pre-processing techniques used as well.

Segmentation of blood vessels and the optic disc plays a significant role in lesion extraction as it reduces the chances of detecting false lesions. Sinthanayothin et al [18] used the intensity of the pixels to determine the areas with high variation to spot the optic disc. A multilayer perceptron neural network was employed for the extraction of blood vessels which involved Principal Component Analysis (PCA) inputs and edge detection. In [19], the authors demonstrated a fully automated system, leveraging recursively growing segmentation algorithms and the Moat operator, to detect HMAs (hemorrhages and microaneurysms) and hard exudates on a relatively small dataset of 30 images. Their system was reported to have sensitivity and specificity scores of 88.5% and 99.7% for exudates, and 77.5% and 88.7% for HMAs. Moccia et al. [20] reviewed various blood vessel segmentation algorithms including methods, datasets and metrics.

Enhancement of retinal images was an important step in every proposed method to increase the efficiency of classification. Extraction of the green channel from the image is seen to be a common approach as it provides more intensity for the vessel pixels, as adopted by Chudzik [21] and Habib [22]. Tan et al. exploited the procedure of converting RGB to LUV color space and removed the varying local contrast and uneven illumination; they proposed two single CNN approaches, one to detect fovea, optic disc and retinal vessel structure in [23], and the other one for HMAs and exudates in [24], the normalization step being the common and pivotal step in pre-processing the images.

Image enhancement techniques such CE (Contrast Enhancement) and CLAHE (Clip Level Adaptive Histogram Equalization) also play an important role in the efficient extraction of lesions. Hajeb et al. [25] demonstrated the use of CLAHE as a pre-processing step in a novel SVM classifier, which took as input both the color fundus images and fundus fluorescein angiograms (FFA), using curvelet transform.

Grinsven et al. [26] demonstrated the efficacy of a CNN using selective data sampling. Interestingly they first preprocessed the images using circular template matching to get the field of view and extracted a square crop, each side having 512 pixels of this circular field of view [27]. Then Contrast Enhancement (CE) was applied, as described in equation 1 below:

$$I_{CE}(x, y; \lambda) = \alpha I(x, y) + \beta G(x, y; \lambda) * I(x, y) + \gamma \quad (1)$$

where, $I$ is the input image, $I_{CE}$ is the contrast-enhanced output, $G(x, y; \lambda)$ represents a Gaussian kernel of scale $\lambda$, $*$ represents the convolution operation. The paper reported $\alpha = 4, \beta = -4, \lambda = 512/30$ and $\gamma = 128$. They demonstrated a significantly faster and more accurate CNN training procedure involving selective sampling (SeS), which dynamically focuses more on training samples that are harder to diagnose.

Khojasteh et al. [28] compared the CE approach suggested in [26] to the frequently used CLAHE and found the model not only gave higher test accuracy but converged faster as well. Khojasteh et al. [29] also demonstrated the efficacy of deep residual network based approaches in exudate detection on the DIARETDB [30, 31] and e-Ophtha [32] databases, improving upon existing approaches, they reported ResNet50+SVM to be significantly better than all the other approaches they tried in the study.

A comprehensive amalgamation of retinal image acquisition, processing and analysis along with clinical relevance of the features has been presented by Abramoff et al. in [33], who also proposed an automated detection [34] of DR strategy that did not miss any diagnoses of NPDR, PDR or ME (macular edema), by combining various detectors for HMAs, exudates etc. The exact architecture of these CNNs were not elaborated in the paper, however, it would be an educated guess to assume the overall system IDx-DR X2.1 would be computationally heavy. Another novel two-stage hierarchical system for classification was proposed in [35], the authors noted that kNN was the best classifier for red lesions as kNN is comparatively robust to imbalanced datasets compared to SVM. [36] proposed an automated DR-screening system based on fundus images, without deep learning.

In [37], a segment based learning approach was presented for DR estimation which, the authors claim, significantly improves the detection performance. Performance analyses of various statistical features through neural networks, SVM, fuzzy logic, classifier fusion etc. were discussed in detail in [38]. Asiri et al. [39] presented a comprehensive survey of various computer-aided diagnostic systems for DR.

## IV. DATASET ANALYSIS

The Kaggle Diabetic Retinopathy Dataset, provided by EyePACS [40] was used in this research work. The competition training set comprised of 35,126 images graded into 5 Diabetic Retinopathy stages with class labels ranging from 0 to 4, and the test set comprised of 53,576 images. The solution file contained the labels for the test images as well. Both the train and test datasets were combined to get a larger dataset of 88702 images, the distribution of which is given below in Fig. 2.

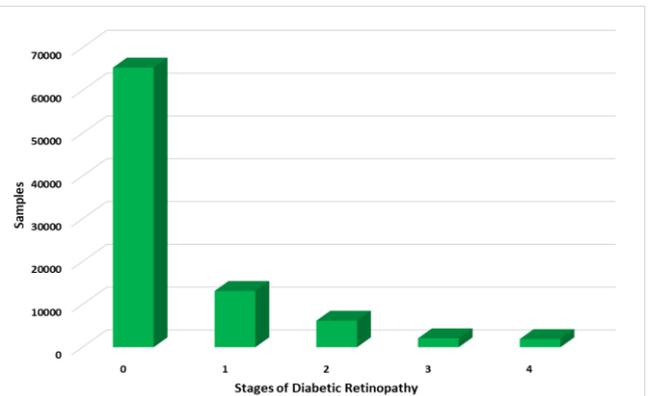

Fig. 2. Original distribution of the combined Kaggle dataset.

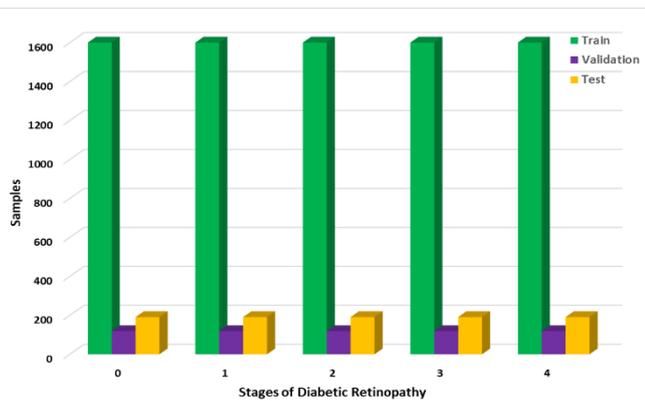

Fig. 3. Balanced distribution of the combined Kaggle dataset.

The weightage per class is observed to be about 73.7%, 14.8%, 6.9%, 2.3% and 2.1% respectively. Due to this high skew in the train set, along with the intention to develop an algorithm that can robustly classify on less data at the edge, only 1910 images per class were selected, which is about 10% of the initial merged dataset. Each of these 1910 images per class were split into 1600 images for training, 119 for validation and 191 for test. Thus, the final set had 8000 images for training, 595 images for validation 955 images for testing the model. This distribution is illustrated in Fig. 3 and henceforth, the word 'dataset' will refer to this dataset.

All the images in the dataset are color fundus images acquired using a variety of cameras and multiple fields of view. Image acquisition specifics like camera model, or details of the field of view were not available. Any patient data like age, history or demographic data also was not provided. Images that were unclear due to various acquisition anomalies like the reflection on the retina, foggy or cloudy images due to lack of proper illumination or focus were dropped (example in Fig. 4) before creating the final dataset, as described above.

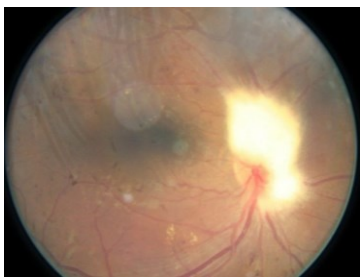

Fig. 4. Image dropped due to strong reflection blurring the optic disc and eyelash shadows in the periphery of the upper hemisphere.

## V. EXPERIMENTAL METHODOLOGY

This section systematically discusses the steps and the research methodology. Firstly, the images were pre-processed to increase the clarity and effectiveness of the CNN model, then augmented and passed into the training pipeline.

### A. Pre-processing

A variety of pre-processing techniques and combinations were experimented with. The pre-processing steps finally chosen have been discussed below in order:

*Choosing the appropriate channel:* The effectiveness of the green channel fundus images in detecting HMAs and exudates has been suggested before in [41-43]. This research corroborates the aforementioned findings and builds up on it to create a fast and efficient classifier. The extracted green channel is used for the next step. Comparison between a grayscale image, traditionally used in CNNs vs an extracted green channel fundus image is shown in Fig. 5.

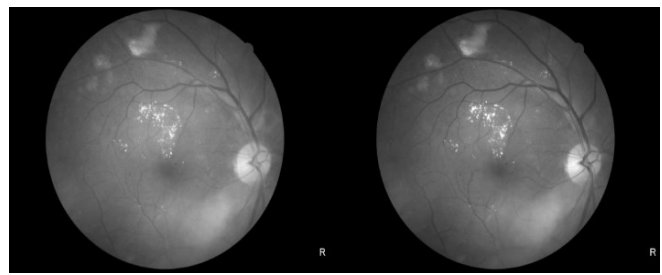

Fig. 5. Grayscale vs Green Channel Fundus Image.

*Image Resizing:* The original images are of various sizes with some images having side lengths of 4000 pixels. To make the training and feature extraction efficient, all images are resized to 256x256. Resizing the images after extracting the green channel and prior to the equalization and clarity boost steps is most computationally efficient.

*Adaptive Equalization:* Adaptive Histogram Equalization is a standard technique used to improve the contrast in image processing, but is prone to over-amplification of contrast in certain parts. CLAHE (Clip Limited Adaptive Histogram Equalization) has been used in this work to get rid of some of the grain and improve contrast. Clip limit (an optimal value of 2 has been chosen) prevents the algorithm from adding extreme features or noise. It operates on small tiles in the image, then merges the results using bilinear interpolation This effect of CLAHE on the green channel image has been demonstrated in Fig. 6.

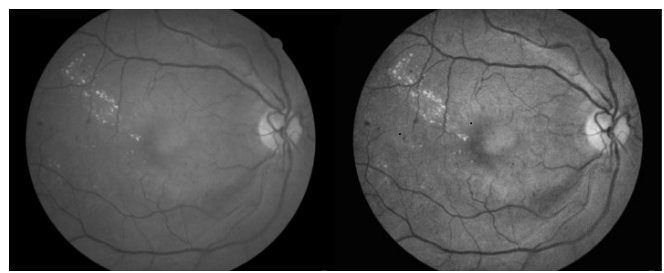

Fig. 6. Green-channel image before and after CLAHE.

*Clarity Boost:* A blurred copy of the above image is created with a blur level of 40. For each pixel p having intensity $X_p$ from the histogram equalization step and the same pixel p in the blurred image having an intensity $Y_p$, $X_p$ is updated according to equation 2 below,

$$X_p = 4X_p - 4X_p + 128 \qquad (2)$$

The improvement is apparent in Fig. 7 below.

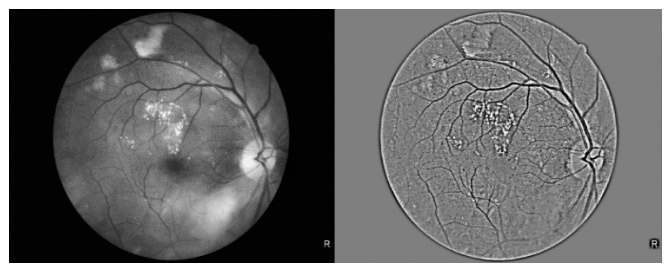

Fig. 7. Features become more prominent after Clarity Boost.

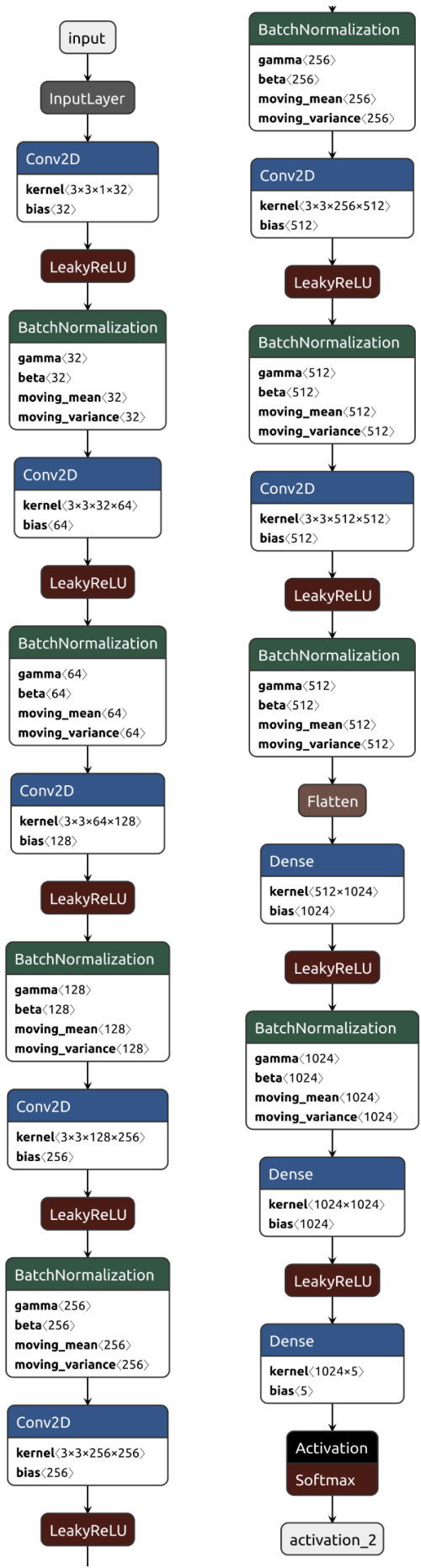

Fig. 8. Architecture of the proposed Convolutional Network.

The images are all of the shape 256x256 and are normalized to the range [0,1] by dividing by 255 to speed up backpropagation. This step is then followed by augmentation.

*B. Augmentation*

These augmentation methods were used in the following order to increase the effective train dataset size:

*Random Flipping:* The images were randomly flipped with a probability of 50% along both axes.

*Random Rotation:* 80% of the images were randomly rotated in the range of $\pm 25°$, to make the model more robust.

*Random Zoom:* The images were randomly zoomed by a random amount within the range [-10%, 10%], with bilinear interpolation and reflective fill mode, reducing edge artifacts.

*C. Model Architecture*

A novel convolutional neural network has been proposed that is capable of classifying the stages of diabetic retinopathy on edge devices, at very high speeds, with great performance. The specifications of the architecture are shown in the Netron Visualization [44] in Fig. 8.

It comprises of 7 Conv-ReLU-BN blocks, where Conv refers to the standard Convolution operation, ReLU refers to the Rectified Linear Activation layer and BN refers to Batch Normalization layer, used to keep the weights in within a threshold reducing internal covariate shift [45, 46], thus stabilizing the training process. Dropout layers [47] were used prior to the last two dense layers, though it hasn't been represented in the Fig. 8 alongside for brevity.

*D. Training Specifications*

The network architecture has already been described above, this section will throw light upon the various version numbers, model hyperparameters, and other specifications used to aid the reproducibility of this work. The network was designed and trained with TensorFlow 2.4.1 using TPUs (Tensor Processing Units) on Google Colab. Weights and Biases [48] was used for experiment tracking.

*ImageDataGenerator* in Keras was used for the input and augmentation pipeline, with a batch size of 32. The callbacks used were *ModelCheckPoint*, that saved only the best model, monitoring accuracy, and *ReduceLRonPlateau* that reduced the learning rate to 10%, if validation loss didn't improve every 5 epochs, no cooldown was used. The third callback was the *WandbCallback*, which is used by Weights and Biases to record the history per epoch, including training and validation loss and accuracy, model save paths etc. The optimizer used was *Adadelta* [49] with an initial learning rate of 1.8, categorical cross-entropy loss and was trained for 200 epochs.

*E. Deployment and Evaluation*

The model was deployed on an OpenMV H7 Plus, which is a microcontroller board powered by STM32H743II ARM Cortex M7 processor with a clock speed of 480 MHz. It has 32MBs SDRAM, 1MB of SRAM along with 32 MB of external flash + 2 MB of internal flash. The maximum framebuffer is about 31 MB. It was deployed through the Edge Impulse integration, streamlining the deployment pipeline.

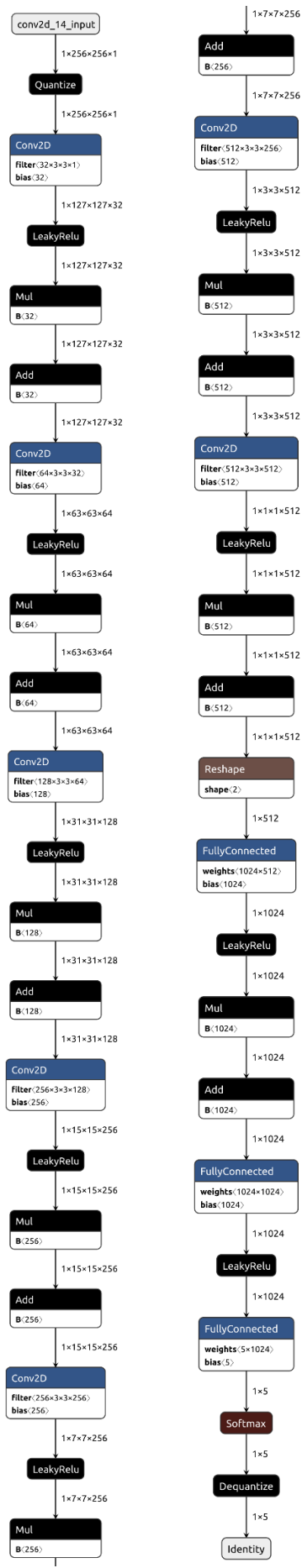

Fig. 9. Schematic of the post-training full-integer quantized CNN.

The model was quantized post-training using full-integer quantization from float32 weights to int8 weights reducing the model size from 70 MB to a meager 5.9 MB, the schematic for the quantized model is shown in Fig. 9. This process uses a representative dataset to measure the dynamic range of the activations and weights, as described in [50, 51].

Performance metrics were calculated using the TF Lite Interpreter and evaluation of speed was done on the microcontroller ARM Cortex M7, the images were passed and the predictions from the quantized TF Lite model were stored to a file, which was then compared against the ground truths.

## VI. RESULTS AND FINDINGS

The model was trained using TensorFlow and Weights and Biases was used for experiment tracking, the progression of accuracy along with the learning rate per epoch can be seen in Fig. 10. Initially, the model makes random predictions and since there are 5 classes the probability of a random prediction being correct is about 20%, hence the initial accuracy. On the $175^{th}$ epoch, a drop in learning rate from 1.8 to 0.18 causes the model to converge faster, and accuracy jumps from 84 to 88%.

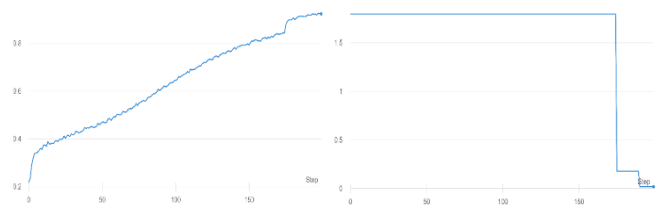

Fig. 10. Loss and Accuracy curves for training.

Confusion matrices and performance metrics were calculated for both the initial and the quantized model, and have been presented in Fig. 11 and Table 1 for the float32 model and in Fig. 12 and Table 2 for the int8 model, giving the reader a detailed understanding of model performance.

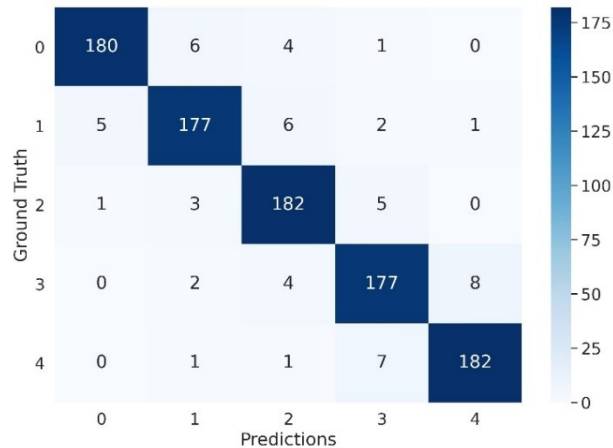

Fig. 11. Confusion matrix for float32 model.

Table 1. Performance metrics of float32 model.

|   | Precision | Recall | F1-score | Support |
|---|---|---|---|---|
| 0 | 0.97 | 0.94 | 0.95 | 191 |
| 1 | 0.94 | 0.93 | 0.93 | 191 |
| 2 | 0.92 | 0.95 | 0.94 | 191 |
| 3 | 0.92 | 0.93 | 0.92 | 191 |
| 4 | 0.95 | 0.95 | 0.95 | 191 |
| Macro-avg | 0.94 | 0.94 | 0.94 | 955 |
| Accuracy |  | 0.94 |  | 955 |

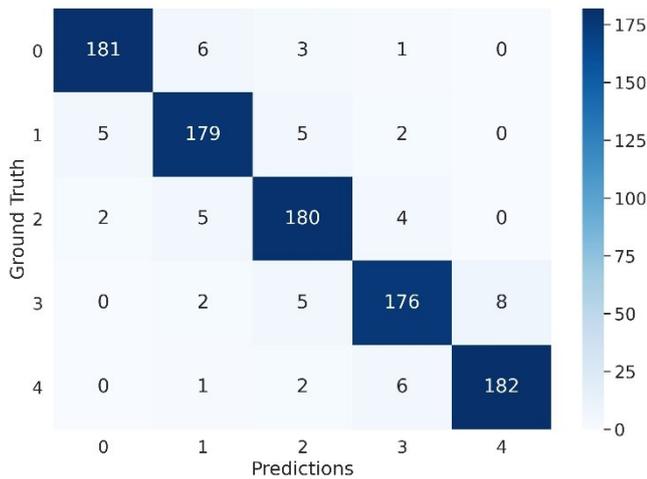

Fig. 12. Confusion matrix for int8 model.

Table 2. Performance metrics of int8 model.

|  | Precision | Recall | F1-score | Support |
|---|---|---|---|---|
| 0 | 0.96 | 0.95 | 0.96 | 191 |
| 1 | 0.93 | 0.94 | 0.93 | 191 |
| 2 | 0.92 | 0.94 | 0.93 | 191 |
| 3 | 0.93 | 0.92 | 0.93 | 191 |
| 4 | 0.96 | 0.95 | 0.96 | 191 |
| Macro-avg | 0.94 | 0.94 | 0.94 | 955 |
| Accuracy |  | 0.94 |  | 955 |

From Figs. 11-12 and Tables 1-2, it is observed that the model has an accuracy of 94% on the held-out test dataset and the quantization step reduces the model size from 70 MB to 5.9 MB with no drop in accuracy, albeit with some negligible variations in the class-wise metrics.

It must be noted that no case of severe NPDR (stage 3) or PDR (stage 4) was predicted as no DR, as that misdiagnosis would have a severely detrimental effect on the automated testing on patients. Any patient with a diagnosis of any level of DR except stage 0 is usually sent for review to a trained ophthalmologist, with varying levels of urgency, which makes this system practical and efficient in classifying DR.

## VII. CONCLUSION AND FUTURE AVENUES

This research successfully demonstrates the ability to classify the stages of Diabetic Retinopathy with a reasonably high level of accuracy of 94%, despite having a tiny model size of 5.9 MB. Leveraging the proposed Diabetic Retinopathy detection network and amalgamating it with state-of-the-art advances in TinyML, this research proves that Computer Vision can be used to diagnose diseases like Diabetic Retinopathy with high speeds and low memory footprints.

Future avenues would involve reducing the model size even further by designing advanced lighter architectures, ablation studies [52] to analyze which neurons can be removed due to low contribution to the classification output thereby reducing the model weight as well as experimenting with low-precision training strategies as discussed in [53-56], and low-precision post-training quantization techniques [57, 58].

This paper demonstrates the viability for Computer Vision systems on the edge to classify Diabetic Retinopathy, and encourages further research in this field.